\documentclass[aps, twocolumn, amsfonts, amsmath, amssymb, superscriptaddress, showkeys]{revtex4-1}

\usepackage{graphicx}

\usepackage{caption,subcaption}
\usepackage[normalem]{ulem}
\usepackage{physics}
\usepackage{tikz,lipsum,lmodern}
\usepackage[export]{adjustbox}
\usepackage{soul}

\usepackage[unicode=true,pdfusetitle,
 bookmarks=true,bookmarksnumbered=false,bookmarksopen=false,
 breaklinks=false,pdfborder={0 0 0},backref=false,colorlinks=true,citecolor=blue,urlcolor=violet 
]{hyperref}
\usepackage{xcolor}
\newcommand{\orcid}[1]{\href{https://orcid.org/#1}{\includegraphics[width=10pt]{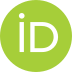}}}

\begin{document}
\title[]{Testing Gravitational Self-interaction via Matter-Wave Interferometry}
\author{Sourav Kesharee Sahoo\orcid{0000-0003-1812-0417}}
\email{sourav.sahoo1490@gmail.com.} 
\affiliation{Department of Physics, BITS-Pilani K K Birla Goa Campus, Goa-403726, India.}
\author{Ashutosh Dash\orcid{0000-0003-2769-085X}}
\email{dashutosh573@gmail.com}
\affiliation{Institute for Theoretical Physics, Goethe University, Frankfurt am Main, Germany}
\author{Radhika Vathsan\orcid{0000-0001-5892-9275}}
\email{radhika@goa.bits-pilani.ac.in} 
\affiliation{Department of Physics, BITS-Pilani K K Birla Goa Campus, Goa-403726, India.}

\author{Tabish Qureshi\orcid{0000-0002-8452-1078}}
\email{tabish@ctp-jamia.res.in}
\affiliation{Center for Theoretical Physics, Jamia Millia Islamia, New Delhi 110025.}
\date{\today}

\begin{abstract}

The Schr\"odinger–Newton equation has frequently been studied as a nonlinear modification of the Schr\"odinger equation incorporating  gravitational self-interaction. However, there is no evidence yet as to whether nature actually behaves this way. This work investigates a possible way to experimentally test gravitational self-interaction. The  effect of self-gravity on interference of massive particles is studied by numerically solving the Schr\"odinger-Newton equation for a particle passing through a double-slit. The results show that the presence of gravitational self-interaction has an effect on the fringe width of the interference that can be tested in matter-wave interferometry experiments. Notably, this approach can distinguish between gravitational self-interaction and environment induced decoherence, as the latter does not affect the fringe width. This result will also provide a way to test if gravity requires to be quantized on the scale of ordinary quantum mechanics.
\end{abstract}
\maketitle

%%%%%%%%%%%%%%%%%%%%%%%%%%%%
\section{Introduction}
%%%%%%%%%%%%%%%%%%%%%%%%%%%%
The emergence of  classicality from quantum theory is an issue which has plagued quantum mechanics right from its inception. Quantum mechanics is linear, and the Schr\"odinger equation allows superposition of any two distinct solutions. However, in our familiar classical world, a superposition of macroscopically distinct states, such as  the state corresponding to two well separated distinct positions of a particle, is never observed \cite{arndt2014testing}. Taking into account  environment induced decoherence \cite{joos2013decoherence,schlosshauer2007decoherence,wang2006quantum}, one may argue that pure superposition states do not survive for long, and the interaction with the environment causes the off-diagonal elements of the \emph{reduced} density matrix of the  system  to vanish. The remaining diagonal terms are then interpreted as classical probabilities.
However, decoherence is based on unitary quantum evolution and if one tried to explain how  a single outcome results for  a particular measurement, one will eventually be forced to resort to some kind of many worlds interpretation \cite{zeh1997}. Another class of approaches to address this issue  invokes some kind of non-linearity in quantum evolution, which may cause macroscopic superposition states to dynamically evolve into one macroscopic distinct state \cite{bassi2013models,carney2021using,carney2022,universe8020058}. Different theories attribute the  origin of the non-linearity to different sources, for instance, an inherent non-linearity in the evolution equation \cite{bassi2003dynamical}, or  gravitational self-interaction \cite{DIOSI1984199,penrose1996gravity,bassi_2022}. Considerable effort has been put into finding ways to test any non-linearity which may lead to the destruction of  superpositions. For example, an experiment in space was proposed, which would involve preparing a macroscopic mirror in a superposition state \cite{marshall2003towards, belenchia2022quantum}. The problem with such experiments, even if they are successfully realized, is that it is difficult to distinguish between the role of decoherence and that of non-linearity in destroying the superposition. An effect that can distinguish between these two possible causes of loss of superposition is sorely needed. This is the issue we wish to address in this work.

In 1984 L. Diosi \cite{DIOSI1984199} introduced a gravitational self-interaction term in the Schr\"{o}dinger equation in order to constrain the spreading of the wave-packet with time. The resulting integro-differential equation, the Schr\"{o}dinger-Newton (S-N) equation, compromised the linearity of quantum mechanics but  provided localized stationary solutions.
It was R. Penrose\cite{penrose1996gravity,penrose2014gravitization} who used the S-N equation to explore the quantum state reduction  phenomenon. He proposed that macroscopic gravity could be the reason for the collapse of the wave function as the wave packet responds to its own gravity.
The effect of gravity and self-gravity on quantum systems have been studied by several authors \cite{colella1975observation,grossardt2016effects,grossardt2016approximations,singh2015possible,yang2013macroscopic,kumar2000single}.

The coupling of classical gravity to a quantum system also addresses the question of whether gravity is fundamentally quantum or classical\cite{carlip2008quantum,Mattingly2005,Wuthrich2005}. This `semiclassical' approach, where gravity is treated in the non-relativistic (Newtonian) limit, provides  simplifications to calculations, but  has faced several theoretical objections\cite{bahrami2014schrodinger}.  However, the ultimate test would be experimental. In such a context, providing an experimental route to test the effect of S-N non-linearity in a simple quantum mechanical context is valuable. 

In the present work, we focus on the evolution of a single isolated massive quantum particle through the non-linear Schr\"{o}dinger-Newton equation. The particle is in a superposition state undergoing a double-slit interference.  Any signature of non-linearity due to gravitational self-interaction in the variation of fringe width with mass should give us an experimental handle on separating the effect of decoherence from gravitational state reduction. 

%%%%%%%%%%%%%%%%%%%%%%%%%%%%%%%%%%%%%%%%%%%%%%%%%%%%%%%%%%%%%%%%%%%%%%%%
\section{The two-slit experiment with self-gravity}

\subsection{Schr\"odinger-Newton equation}
The S-N equation originated from the context of semi-classical gravity, first introduced by M\"oller \cite{moller1962theories} and Rosenfeld \cite{rosenfeld1963quantization} independently. The fundamental interaction considered in this approach is the coupling of quantized matter with the classical gravitational field \cite{mattingly2005quantum,kibble1981semi,bahrami2014schrodinger}. In this approach, the Einstein field equations get modified as, 
\begin{equation}
          R_{\mu \nu} + \frac{1}{2} g_{\mu \nu} R = \frac{8 \pi G}{c^4} \langle \Psi | \hat{T}_{\mu \nu}| \Psi \rangle \label{e2}
          \end{equation}
where the term on the right hand side is the expectation value of the energy-momentum tensor with respect to the quantum state $|\Psi\rangle$ of matter.  
This semi-classical modification has been studied with reference to the necessity of quantizing gravity \cite{eppley1977necessity,page1981indirect}. 
The prescribed modification to the Einstein  field equation leads to the Schr\"odinger-Newton equation \cite{bahrami2014schrodinger,van2011schrodinger,salzman2005investigation,giulini2012schrodinger},

  \begin{equation}
          \Bigg[ - \frac{\hslash^2}{2 m} \nabla^2 - G m^2 \int \frac{|\Psi (r',t)|^2}{|r - r'|} d^3 r'\Bigg] \Psi(r,t) = i \hslash \frac{\partial \Psi (r,t)}{\partial t}. \label{e1}
      \end{equation}
 The above equation can be seen as a \emph{non-linear} modification
of the Schr\"odinger equation. The non-linearity breaks the unitarity of the quantum dynamical
evolution, and opens up the possibility of a dynamical reduction of the
wavefunction, generally referred to as collapse.  It is then not surprising that such modification to  linear quantum mechanics and classical gravity invites criticism \cite{anastopoulos2014problems}. Apart from this, there have
been several  other  collapse models that have been investigated in the
literature \cite{bassi2013models,ghirardi1990markov,adler2007collapse}.
 However this approach has to be tested both theoretically and experimentally, if one wants to rule it out. Our approach is to check whether it has any significance in the emergence of classicality at all, more so if there is an effect that can be experimentally tested. In future, if S-N equation gets ruled out by experiments then the particular coupling considered in equation~(\ref{e2}) will  also get ruled out and other types of coupling between gravity and matter fields could be considered \cite{grossardt2017newtonian}. 

\begin{figure}
    \centering
    \includegraphics[width=7cm]{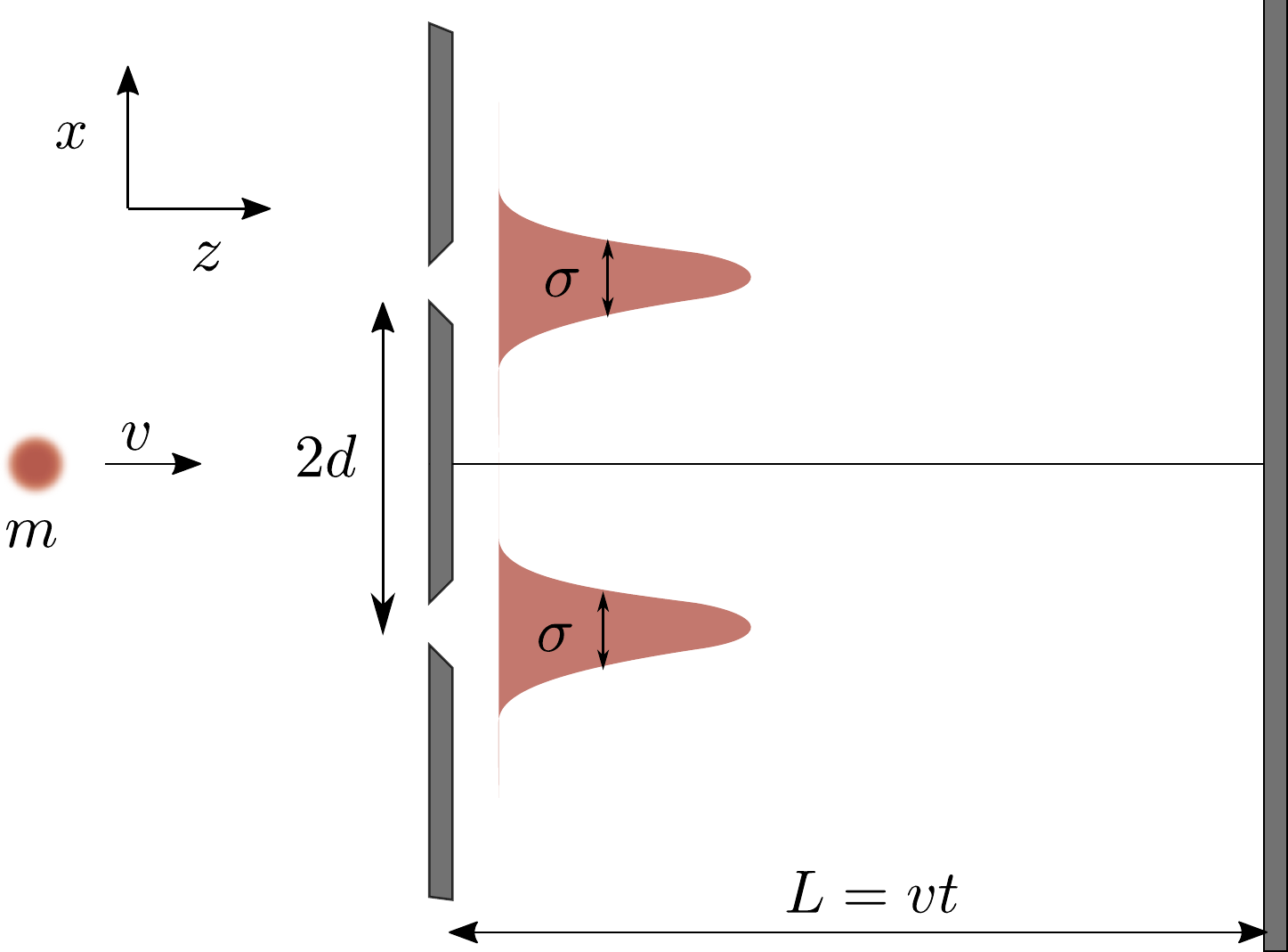}
    \caption{Schematic diagram of two-slit interferometer for a massive particle}
    \label{fig:twoslit}
\end{figure}

We start by making the S-N equation dimensionless,  using  scaling parameters: 
\mbox{$\tilde{r}=r/\sigma_r,$}
\mbox{$\tilde{m}=m/m_r~,$}
\mbox{$\tilde{t}=t/t_r~.$}
The length scale factor $\sigma_r$ is determined by the natural length scale of the problem.  Once the length scale factor $\sigma_r$ is fixed, for instance by experimental considerations (which we discuss in the subsequent section), the other scale factors are determined in terms of $\sigma_r$ and natural constants:
\begin{equation}
t_r=\left(\frac{\sigma_r^5}{G \hslash}\right)^{\frac{1}{3}},~~~
m_r=\left(\frac{\hslash^2}{G \, {\sigma_r}}\right)^{\frac{1}{3}}.
\label{eq:scalefactors}
\end{equation}
The re-scaled, dimensionless equation is
\begin{equation}
          \Bigg[ - \frac{\tilde{\nabla}^2}{2 \tilde{m}}  - \tilde{m}^2  \iiint  \frac{|\tilde{\Psi} (\tilde{r}',\tilde{t})|^2}{|\tilde{r}-\tilde{r}'|} d^3\tilde{r}'\Bigg] \tilde{\Psi}(\tilde{r},\tilde{t}) = i  \frac{\partial \tilde{\Psi} (\tilde{r}',\tilde{t})}{\partial \tilde{t}}, \label{s2}    \end{equation}  
where, $\tilde{\Psi}\left(\tilde{r},\tilde{t}\right)=\sigma_r^{-\frac{3}{2}}\,\Psi\left(r,t\right)$.
The problem now  has only one scaling parameter $\tilde{m}$ which is dependent on $m_r$. 

\subsection{Formulation of the problem}

We analyze the effect of self gravity on the interference produced by a  particle of mass $m$ passing through a two-slit interferometer (Fig. \ref{fig:twoslit}). The two slits are separated by a distance $2d$ along the $x$-axis. The particle is assumed to travel along $z$-axis towards the screen with a constant velocity $v$. 

As the particle emerges from the two-slit, we assume that the initial state is a superposition of two Gaussian wave-packets. 
For the purpose of interference, the dynamics along the $z$-axis is unimportant. It only serves to transport the particle from the slits to the screen by a distance $L=vt$ in a fixed time $t$. The interference results only from the spread and overlap of the wave-packets in the $x$-direction. Hence we assume the initial wave-function to be
spread along the $x$-direction alone. For calculational simplicity, we assume no spread along the other two directions:
\begin{equation}
          \tilde{\Psi}(\tilde{x},0)=A\left[e^{\frac{-(x-d)^2}{2\, \sigma^2}}+e^{\frac{-(x+d)^2}{2\, \sigma^2}}\right],
      \end{equation} 
where $\sigma$ is the width of  each Gaussian. We completely ignore the time evolution in the $y$ or $z$-directions.

Since we start with a wave-function restricted to the $x$-axis, the potential due to self-gravity in equation~(\ref{s2}) becomes
 \begin{equation}V_G = -\tilde{m}^2\iiint\tfrac{|\tilde{\Psi} (\tilde{x}',\tilde{t})|^2\,\delta (\tilde{y}-\tilde{y}')\, \delta (\tilde{z}-\tilde{z}')}{\sqrt{(\tilde{x}-\tilde{x}')^2+(\tilde{y}-\tilde{y}')^2+(\tilde{y}-\tilde{y}')^2}}
           d\tilde{x}' \, d\tilde{y}' \,d\tilde{z}'. 
\label{eq:potential}
           \end{equation}
Performing the  delta-function integral, equation (\ref{s2}) reduces to an effective 1-d equation
\begin{equation}
          \Bigg[ - \frac{1}{2 \tilde{m}} \frac{\partial^2}{\partial \tilde{x}^2} - \tilde{m}^2  \int \frac{|\tilde{\Psi} (\tilde{x}',\tilde{t})|^2}{|\tilde{x}-\tilde{x}'|} d\tilde{x}'\Bigg] \tilde{\Psi}(\tilde{x},\tilde{t}) = i  \frac{\partial \tilde{\Psi} (\tilde{x},\tilde{t})}{\partial \tilde{t}} \label{s4}.
      \end{equation}
The Schr\"odinger-Newton equation~(\ref{e1}) is a non-linear integro-differential equation and is hard to solve analytically. We could use perturbative approximations, but in order to understand  the effect of self-gravity on interference phenomena, approximation methods will not be helpful.
We therefore resort to numerical solution. 

Now a massive particle is expected to
lose coherence during the time evolution and it is obvious that there will also be 
decoherence effect due to gravitational and other kinds of interaction with the
environment. This may lead to  suppression of interference  in a matter-wave
interferometry experiment. For large mass values, one cannot
confidently attribute this loss of interference to self-gravity, since environment
induced decoherence also leads to exactly the same effect
\cite{qureshi2008decoherence,zeilinger2003}.
The purpose of this work is to separate out the
effects of self-gravitational interaction from those of decoherence. We therefore consider a pure state superposition state evolving only under  self-gravitational potential.

\newcommand\image{\adjustbox{valign=m,vspace=1pt}{\includegraphics[width=0.38\linewidth]{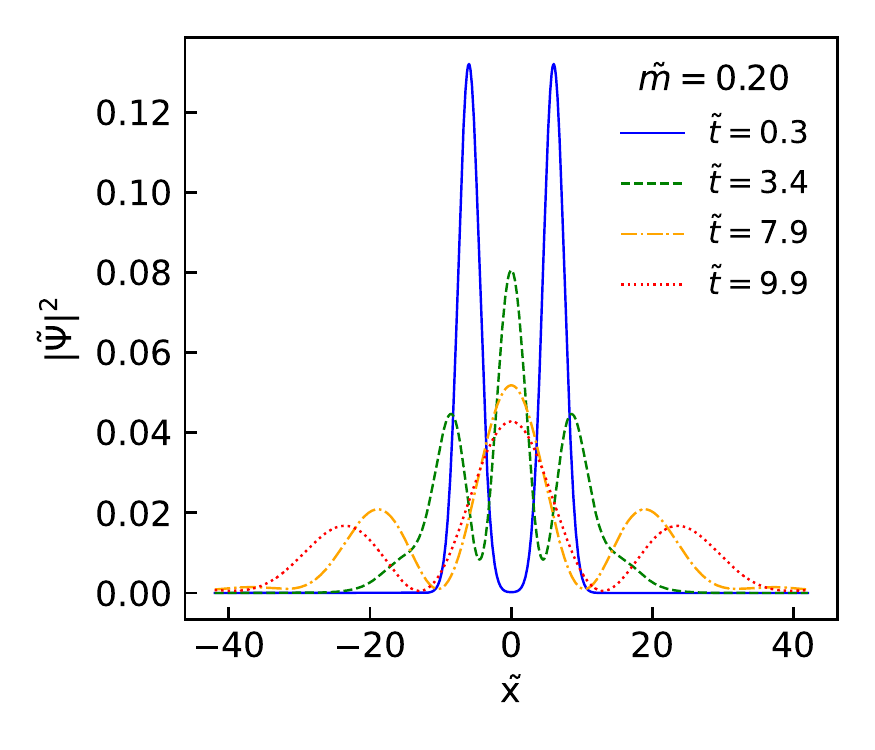}}}
\newcommand\aimage{\adjustbox{valign=m,vspace=1pt}{\includegraphics[width=0.38\linewidth]{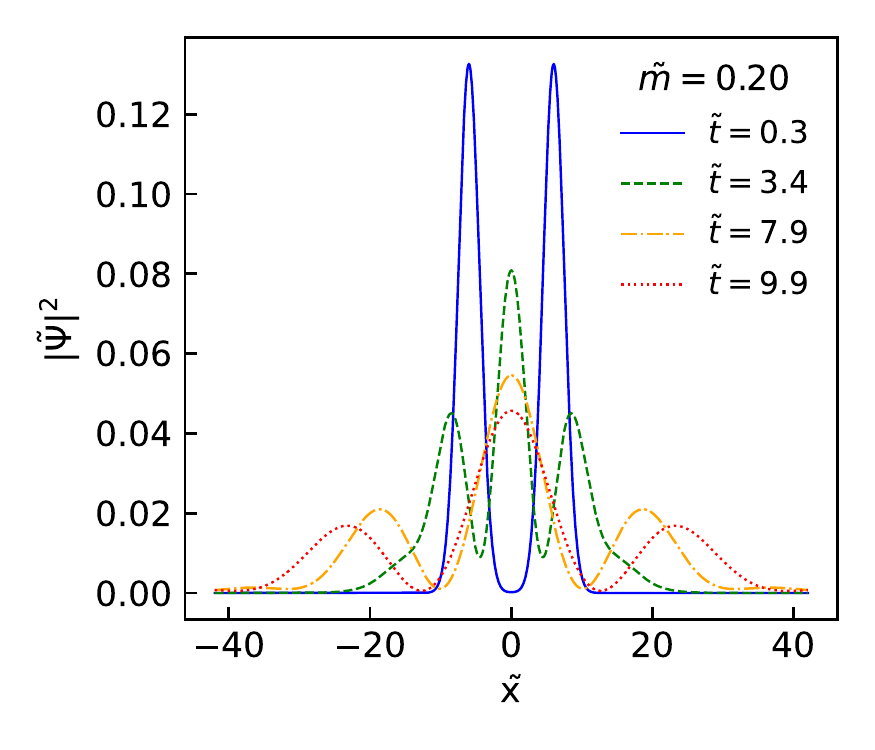}}}
\newcommand\animage{\adjustbox{valign=m,vspace=1pt}{\includegraphics[width=0.38\linewidth]{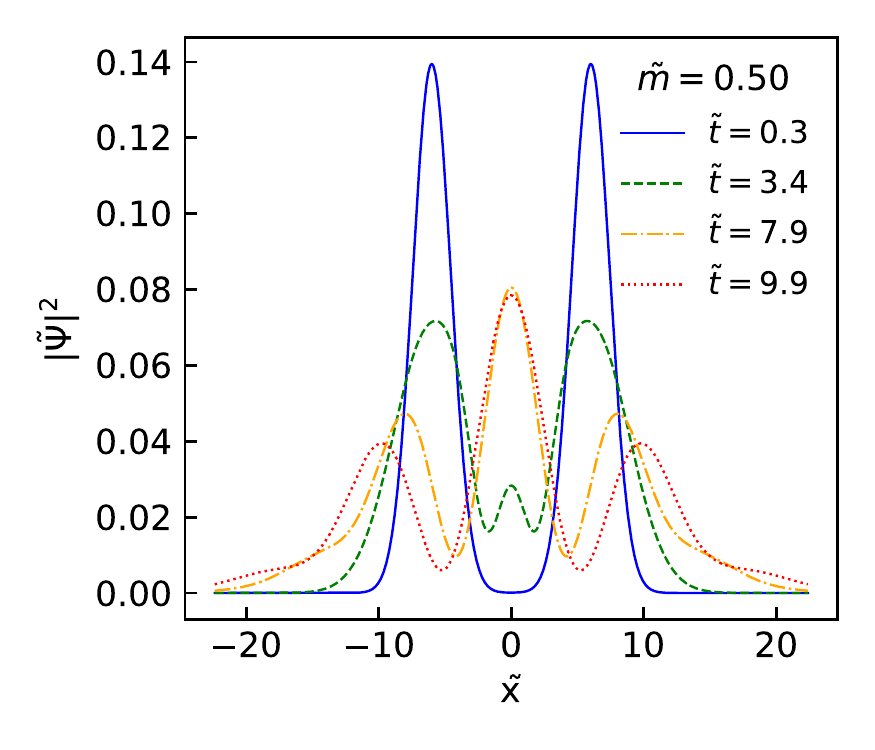}}}
\newcommand\annimage{\adjustbox{valign=m,vspace=1pt}{\includegraphics[width=0.38\linewidth]{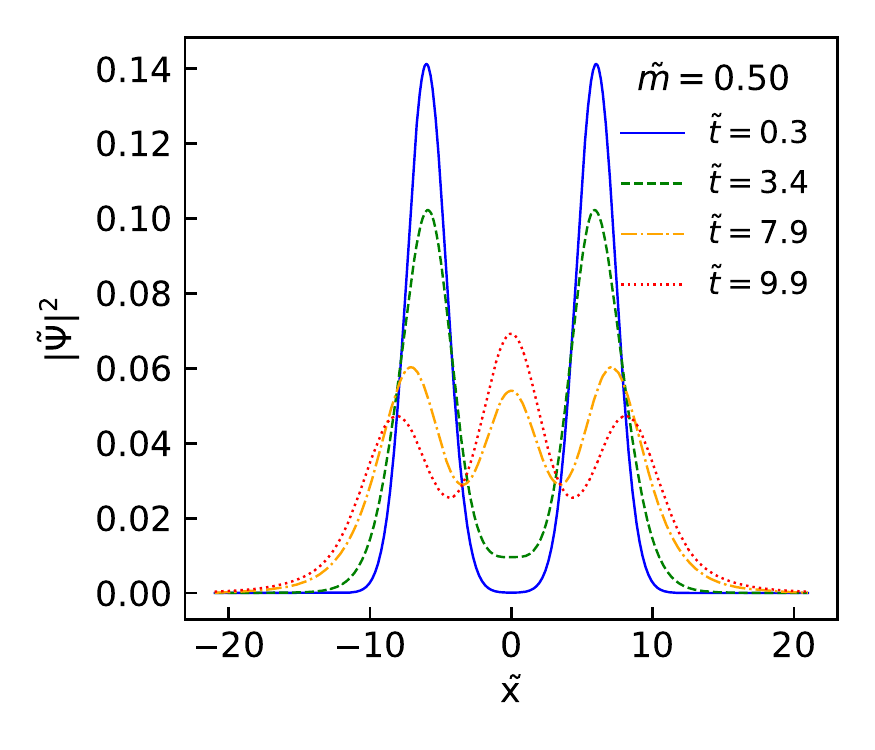}}}
\newcommand\annnimage{\adjustbox{valign=m,vspace=1pt}{\includegraphics[width=0.38\linewidth]{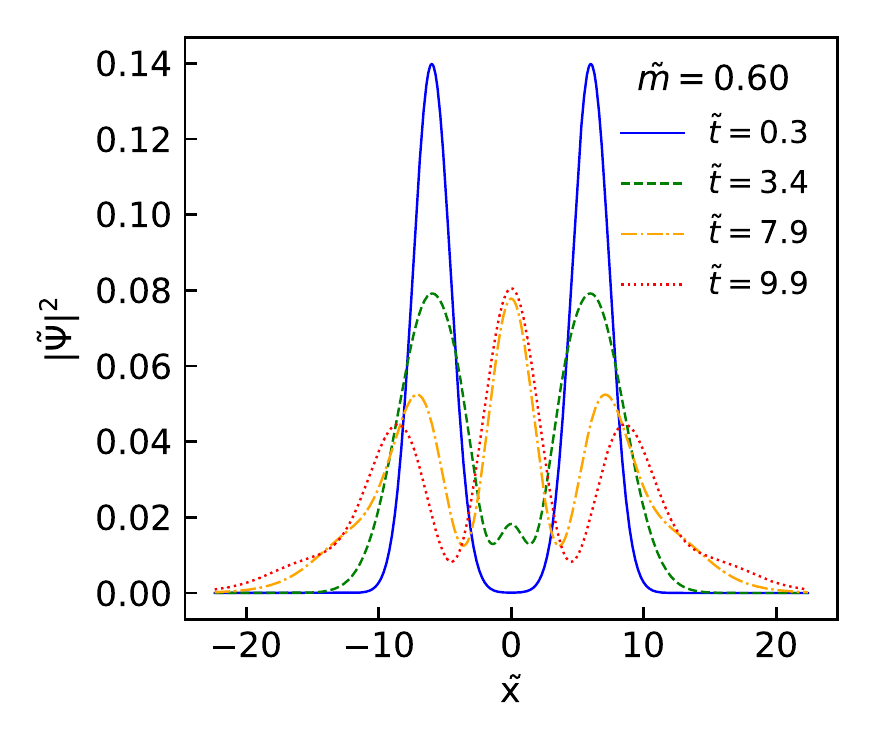}}}
\newcommand\annnnimage{\adjustbox{valign=m,vspace=1pt}{\includegraphics[width=0.38\linewidth]{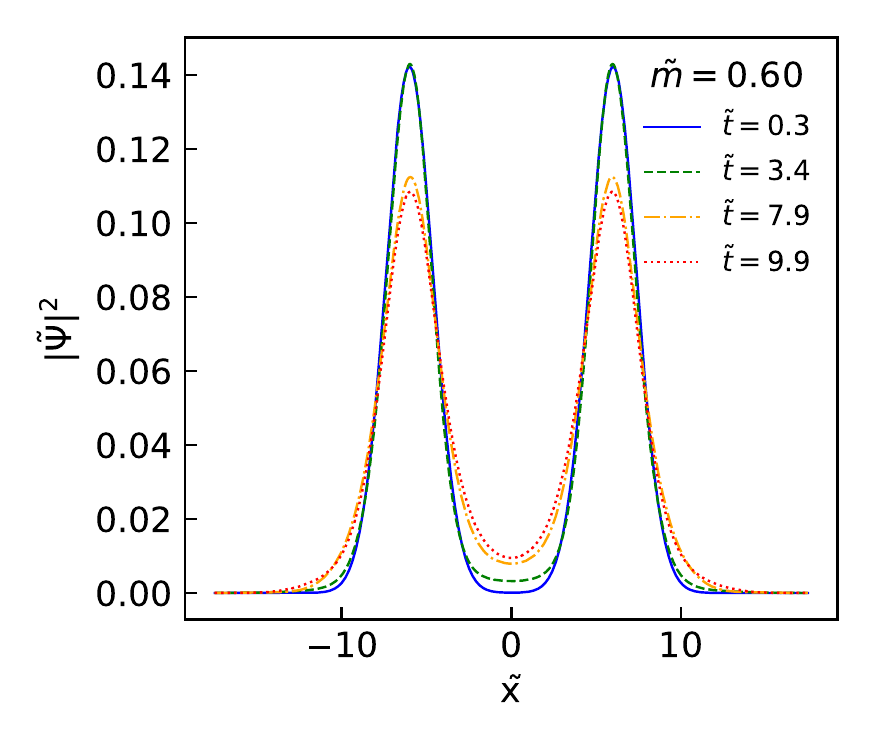}}}

\begin{figure*}[t]
\centering
\begin{tabular}{ccc}
       & ~~Free Schr\"{o}dinger evolution & S-N evolution \\
$ \tilde{m}=0.20$  & \image & \aimage  \\
$ \tilde{m}=0.50$  & \animage & \annimage  \\
$ \tilde{m}=0.60$  & \annnimage & \annnnimage  
\end{tabular}
\caption{Comparison of onset of quantum interference as the superposition evolves with time for different mass values with and without self-gravity.}
\label{Fig:fringes}
\end{figure*}

\section{Numerical results and Discussion}

\subsection{The Numerics}
We solve  equation (\ref{s4}) numerically to obtain the solution $\tilde{\Psi}(\tilde{x},\tilde{t})$ for all rescaled time $\tilde{t}$. We have used Crank-Nicolson method \cite{Crank_1947, smith1985numerical, fletcher2012computational}, as it preserves unitarity at each time step. 

We have used  $d=6 \, \sigma_r$ and $\sigma=2 \, \sigma_r$.
The spatial extent is $[-70,70]$, which is divided into $2000$ spatial grid points and the temporal grid length is taken from $0$ to $10$ and is divided into $1000$ time steps. Hence, $\delta\,\tilde{x}=0.07$ and $\delta\,\tilde{t}=0.01$. 
For the Crank-Nicolson method,  the Courant-Friedrichs-Lewi ($\text{CFL}$) condition necessary for convergence, is satisfied since $\frac{\delta\,\tilde{t}}{\delta\,\tilde{x}} \sim 0.01 <1$.

The boundary points $-70, 70$ actually represent  numerical infinity. However,  as the wave-function evolves in time, the quantum mechanical spread could cause the solution to reach the numerical boundary. Once it reaches the boundary,  the evolution in the next time step causes $\Psi$ to reflect back and affects the entire solution. To avoid this undesirable effect, we have taken the boundary large enough such that the evolved wave-packets do not reach the boundary within the time of evolution considered. 

In order to avoid the singularity in the 1-D form of the  self-gravity potential (equation  (\ref{eq:potential}), we used a regularized form of the potential $V_G(\tilde{x}) = - \tilde{m}^2  \int \frac{|\tilde{\Psi} (\tilde{x}',\tilde{t})|^2}{\sqrt{(\tilde{x}-\tilde{x}')^2+\epsilon^2}} d\tilde{x}'$, where $\epsilon$ is a small dimensionless parameter. In the limit $\epsilon\to 0$ one recovers the original potential. %This can be considered as the gravitational equivalent of the soft-core coulomb potentials used in atomic physics. 
The value of $\epsilon$ is fixed at $0.01$.

\subsection{The Interference}
The interference patterns for different values of $\tilde{m}$ are plotted in figure ~(\ref{Fig:fringes}). The  $x$-axis is position in units of $\sigma_r$  and the $y$-axis is the (dimensionless) probability density $|\tilde{\Psi}(\tilde{x},\tilde{t})|^2$. As one moves from mass $\tilde{m}=0.20$ to $\tilde{m}=0.60$, the crossover from temporal emergence of interference to complete suppression of it, due to the effect of self-gravity, is beautifully brought out. At intermediate values of mass the interference is seen with lower visibility. In contrast, in the absence of self-gravity, interference is seen even at large mass values.

In the usual two-slit interference scenario, the fringe width is equal to $\lambda L/2d$, where $\lambda$ is the de Broglie wavelength of the particle, $2d$ the slit separation, and $L$ the distance between the double-slit and the screen. For a particle of mass $m$ traveling with a velocity $v$, the de Broglie wavelength is $\lambda = h/mv$. Taking the distance traveled by the particle as $L=vt$, the fringe width turns out to be $w=ht/2md$. Thus, for a fixed $t$ the fringe width varies inversely with the mass of the particle. Even if the particle experiences environment induced decoherence, although the interference visibility goes down, the fringe width remains unaffected \cite{qureshi2008decoherence}. Therefore, any deviation of the fringe width from $1/m$ dependence should be a signature of the effect of self-gravity.

\begin{figure*}[t]
     \centering
     \begin{subfigure}[b]{0.45\textwidth}
         \centering
         \includegraphics[width=\columnwidth]{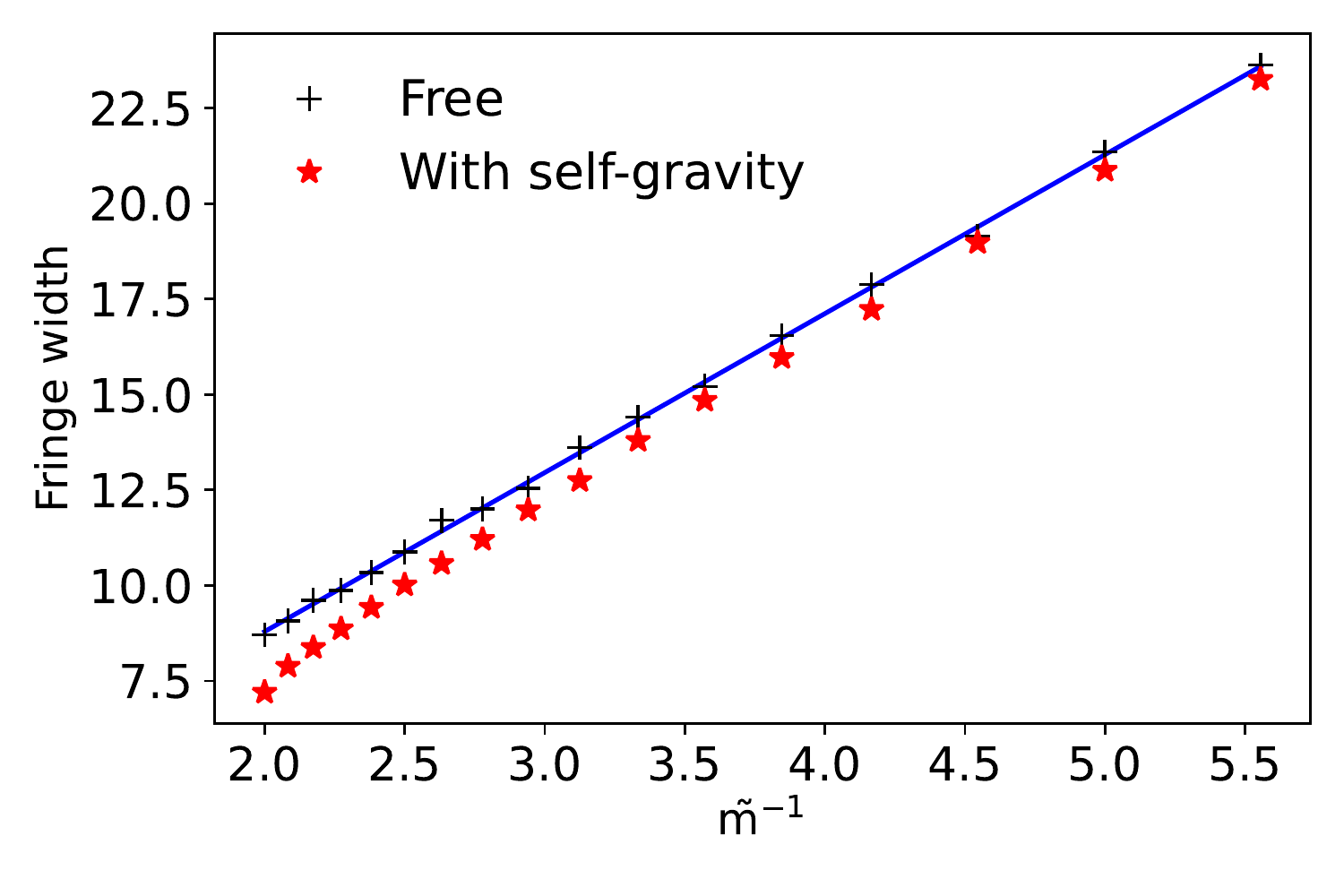}
         \caption{For the full mass range}
    \end{subfigure}
  \begin{subfigure}[b]{0.45\textwidth}
         \centering
          \includegraphics[width=\columnwidth]{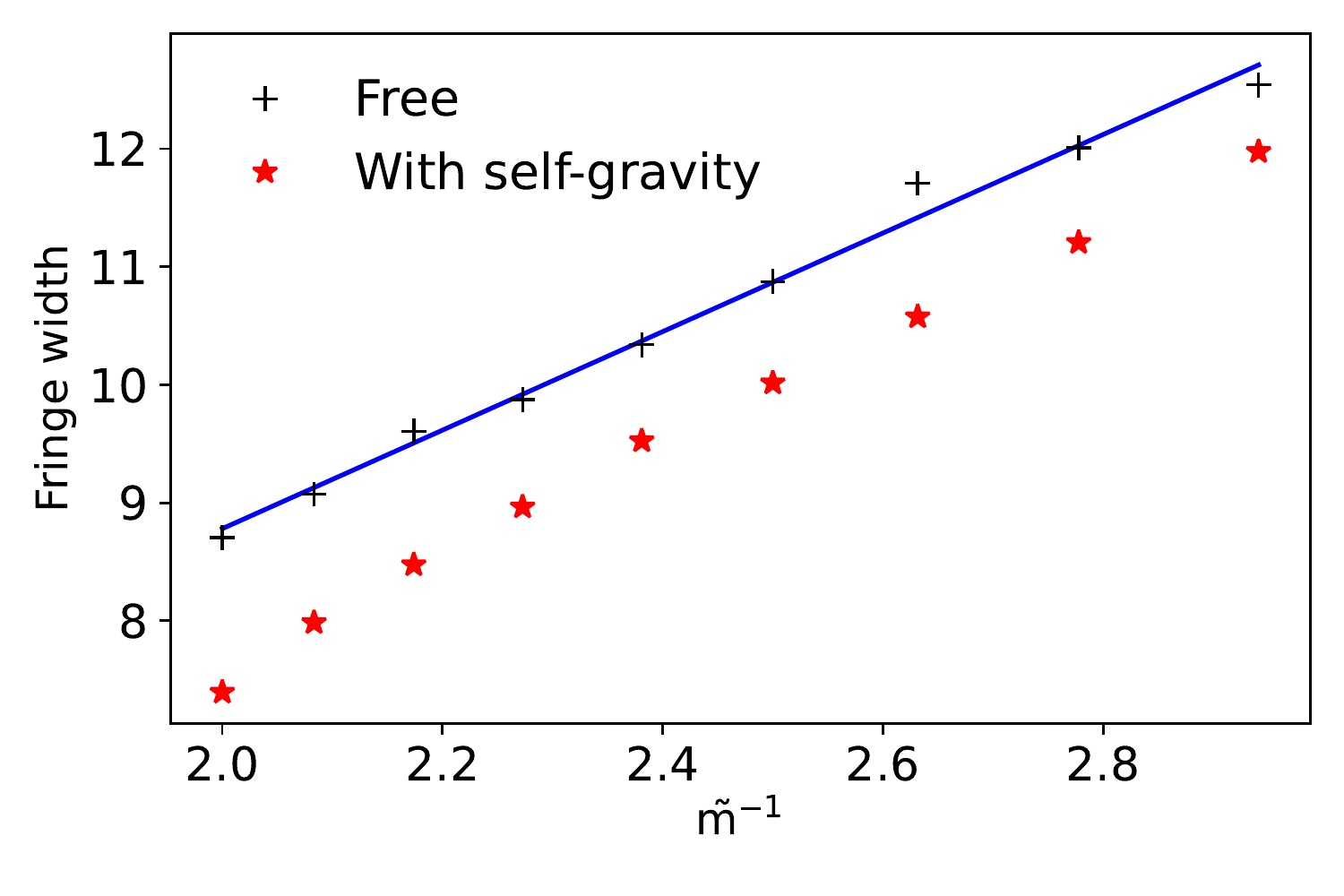}
          \caption{Zoomed in to high $\tilde{m}$ values}
          \label{fig:zoomin}
     \end{subfigure}
      \caption{Fringe width  $w$ (in units of $\sigma_r$) from simulated evolution  as a function of $1/ \tilde{m}$ at time $\tilde{t}=8.9$. The $+$ symbols represent $w$ without  self-gravity, the straight line through them being the trend line; the red stars represent $w$ in the presence of self-gravity. For larger mass, in the presence of self-gravity, the deviation of $w$ from $1/ \tilde{m}$ behavior is more evident.}
      \label{fig:fringewidth}
     \end{figure*}

The fringe width $w$ is calculated from the simulated results as follows. It is assumed that a central peak in the probability distribution is a necessary signature of interference. We calculate $w$ as the distance between the central peak and its nearest interference maximum. One can see from  figure~(\ref{fig:interference}) that the interference peaks are well defined at  $\tilde{t}=8.9$, for various values of $\tilde{m}$. Thus, without  ambiguity,  we calculate  the fringe width from the probability distribution for varying $\tilde{m}$, both with and without the self-gravity  potential term.
We plot  $w$ vs $1/ \tilde{m}$  with and without self-gravity in figure~(\ref{fig:fringewidth}).
The results clearly show that in the presence of the self-gravity potential, $w$ deviates from $1/ \tilde{m}$ dependence as the mass of the particle increases. We believe this should form a clear test of gravitational self-interaction.

 We also notice that as the mass increases, the spread in the wave-function is suppressed by the self-gravity effect. This is clearly seen in Fig. \ref{fig:interference}, where we plot the probability density  at $\tilde{t}=8.9$ for different mass values. For smaller masses, the wave-function spreads enough so that the two wave-packets overlap to result in interference. For much larger masses the gravitational self-interaction suppresses the spread of the wave-packets so that they are not able to overlap and do not lead to any interference. This behavior is consistent with the original aim of introducing the S-N equation.
 
 However, we would also like to point out that we do not see an ``attraction" between the two wave-packets for the mass ranges and time duration considered here. This too is a desirable feature if the  S-N equation is to potentially demonstrate any collapse of the wave-function. In real experiments the Schr\"odinger Cat states, i.e., the states in a spatially separated superposition of disjoint localized states, can only collapse to one of the two separated parts, never in the middle of the two.

 \subsection{``Attraction" between peaks}
\begin{figure}
    \centering
    \includegraphics[width=\columnwidth]{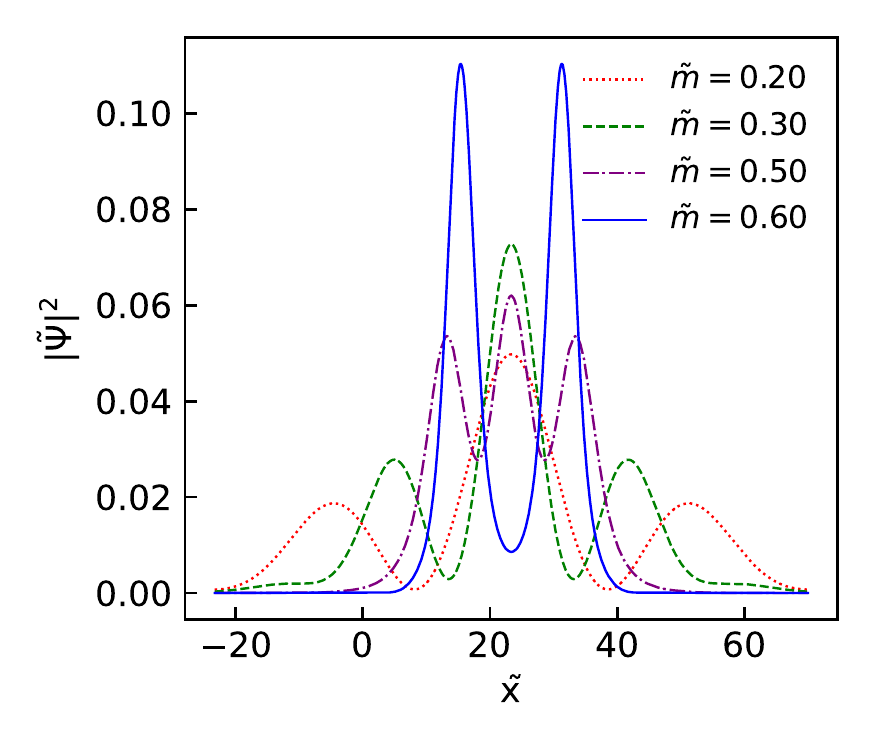}
    \caption{Interference pattern at time  $ \tilde{t}=8.9$ for different values of $\tilde{m}$. The interference gets progressively unsharp as the mass increases, until it is completely suppressed.}
    \label{fig:interference}
\end{figure}

 It is generally expected that if the wave-function has  two lobes, the self-gravitational interaction will lead to an ``attraction" between the two, in the sense that dynamical evolution will bring them closer together. In Figure~ (\ref{fig:interference}), there is apparently no noticeable attraction, within the time range considered here. %This calls for some explanation. We investigated this aspect in somewhat greater detail, and  now understand the reason for the  apparent absence of attraction. 
 
\begin{figure*}[t]
    \centering
     \begin{subfigure}[b]{0.45\textwidth}
         \centering
    \includegraphics[width=\columnwidth ]{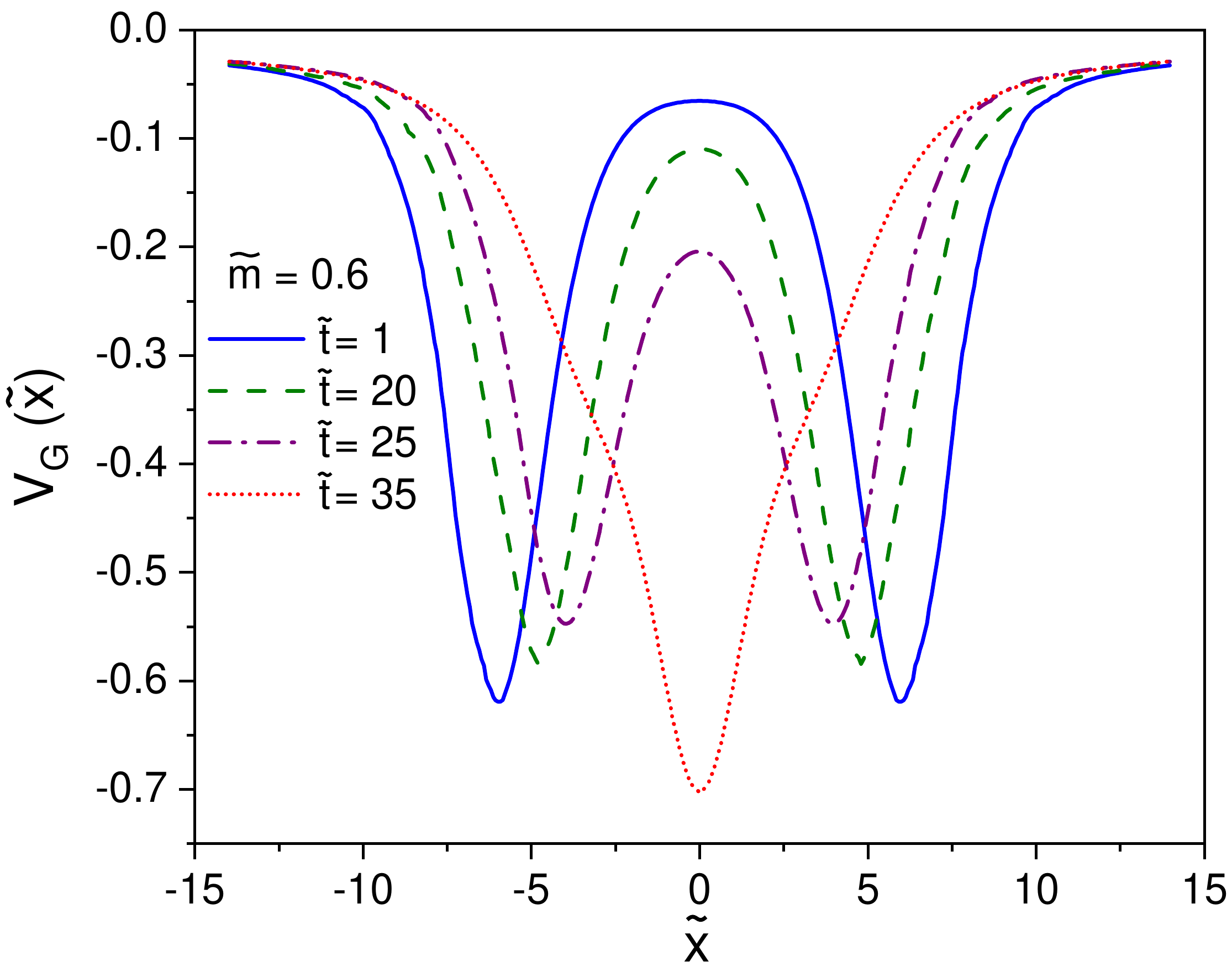}
    \caption{$\tilde{m}=0.60$}
    \label{fig:pot-0.6}
    \end{subfigure}
    \begin{subfigure}[b]{0.45\textwidth}
         \centering
    \includegraphics[width=\columnwidth ]{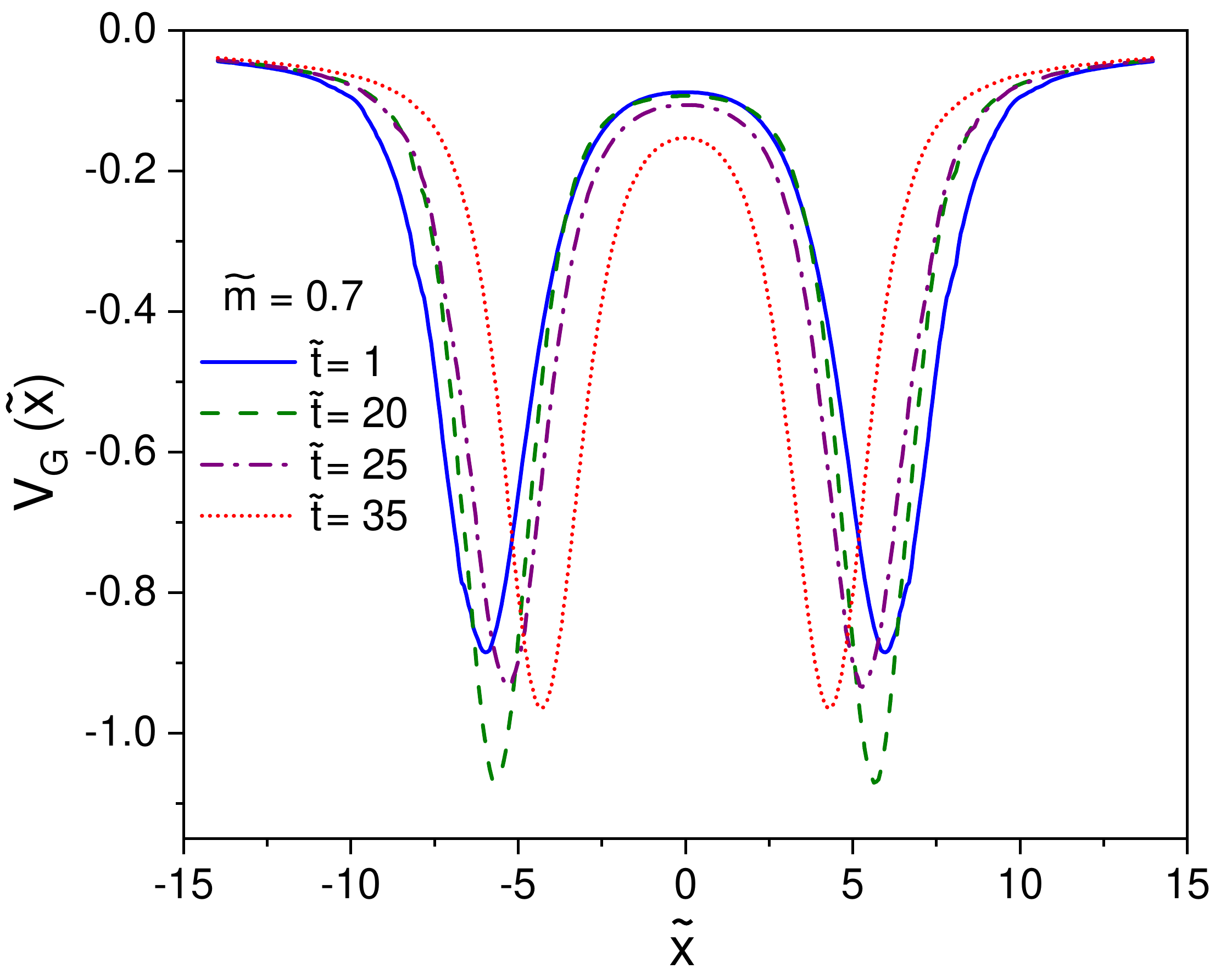}
    \caption{$\tilde{m}=0.70$}
    \label{fig:pot-0.7}
    \end{subfigure}
    \caption{The self-gravity potential $V_G(\tilde{x})$ for higher masses, plotted at various times to show its time evolution.}
    \label{fig:pot}
\end{figure*}

 We take a closer look at the form of self-gravity potential as time evolves, for a much longer time range (figure~\ref{fig:pot}).
 The initial wave-function consists of two disjoint lobes and hence the potential  peaks  near the centers of the two wave-packets. The effect of this is seen as a narrowing of the two wave-packets about their centers.  As  time evolves,  there is a competition between two effects: the narrowing of each wave-packet due to self-gravity, and the broadening effect of Schr\"odinger evolution. 
 
 For low enough masses, the broadening effect of quantum evolution seems to dominate, the wave-packets overlap and interference is observed. For higher masses, apart from the narrowing effect due to the dominance of self-gravity, there is also overlap of the wave-packets  at long times. This contributes to the potential in the region between the two peaks and results in the peaks in the potential drawing closer together until eventually there is a single central peak. The effect of this is that the two wave-packets appear to ``attract" each other, until eventually there is a single central peak.

 Na\"ively one would have expected that the attraction between the peaks would be stronger as the mass increased. However, for reasons described above, the higher the mass, the slower is the attraction between the peaks.

\begin{figure*}[t]
    \centering
    \begin{subfigure}[b]{0.45\textwidth}
         \centering
    \includegraphics[width=\columnwidth ]{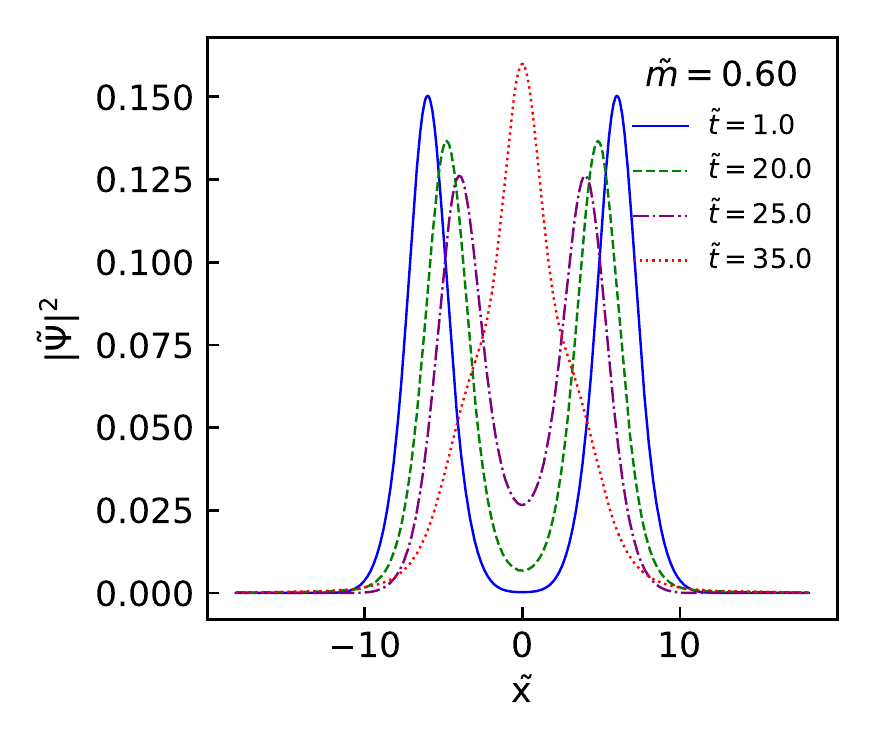}
    \caption{$\tilde{m}=0.60$}
    \end{subfigure}
    \begin{subfigure}[b]{0.45\textwidth}
         \centering
    \includegraphics[width=\columnwidth ]{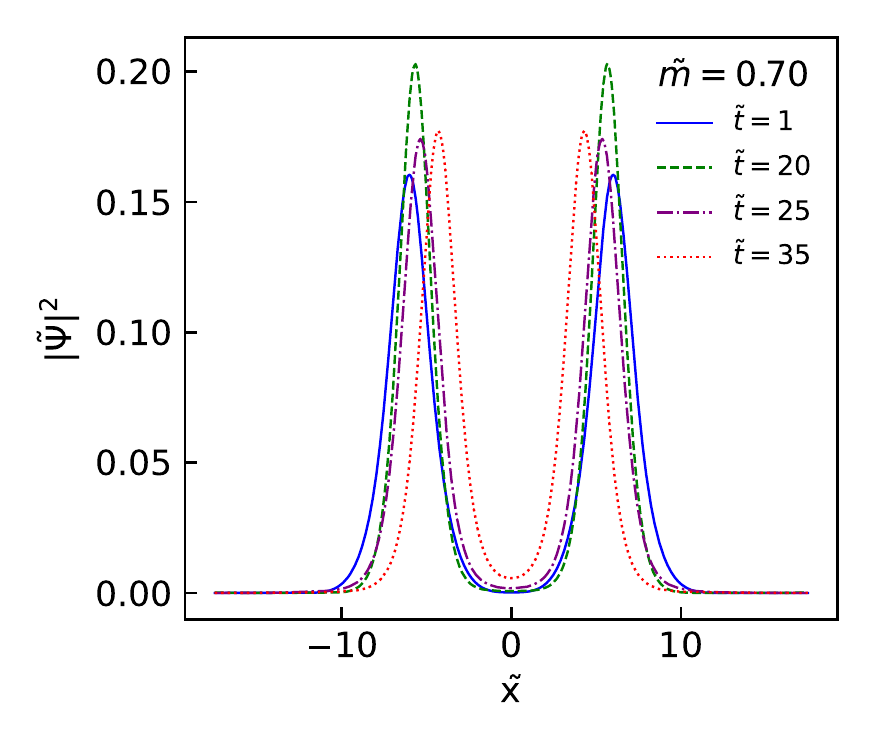}
    \caption{$\tilde{m}=0.70$}
    \end{subfigure}
    \caption{Probability distribution for relatively large mass values, showing  attraction due to self-gravitational effects, and finally merge into a single peak.}
    \label{fig:prob0.80}
\end{figure*}

\subsection{Experimental feasibility}\label{feasibility}
Lastly we would like to discuss what kind of challenges our proposal poses for the experiments, if one were to try observing this effect in some experiment. As seen from Fig. \ref{Fig:fringes} and Fig. \ref{fig:fringewidth}, the effect of self-gravity on the fringe-width is visible for $\tilde{m}\sim 0.5$ for $\tilde{t} \sim 8$.
If we consider $\sigma_r = 1.112$ nm, it leads us to $m_r=31.94\times 10^9$ u and $t_r=0.623$ s, which means that for particles of mass about $16\times 10^9$ u, the self-gravity effect should be observable after about $5$ seconds of time evolution. 
The slit separation required will be about 13 nm. Now interferometry with large molecules has shown a steady progress, with the interference of $C_{70}$ fullerene molecules through Talbot-Lau interferometer being a prominent example \cite{brezger2002matter}. Probably the best technology at present is the  optical time-domain ionizing matter-wave (OTIMA) interferometer \cite{Arndt_2013}. Vienna Kapitza–Dirac–Talbot–Lau interferometer is another one that is capable of using such high mass range, approximately 6509 u \cite{gerlich2007kapitza, rodewald2018isotope}. It is hoped that in the future, particles of mass $10^8$ u, like gold clusters, might be diffracted with the OTIMA scheme \cite{arndt2014testing}. However, even this mass range is too small for observing the effect due to self-gravity. This is exemplified by the fact that if one insists on looking for self-gravity effects for particles of mass  $10^8$ u, one would need times of ridiculous magnitude, of the order of $10^{10}$ s, to see the self-gravity effects. 

So, the message is that one would need to study interference of particles of mass of the order of $10^{10}$ u if one hopes to see any signature of self-gravitational interaction. This looks challenging with the state of the art technology. However, the advantage of our approach to testing self-gravity is that one need not go for creating macroscopic superposition states, the so-called Schr\"odinger cat states. One just needs to do an interference experiment, which should be simpler than creating Schr\"odinger cat states.

\section{Conclusion}

In conclusion, we find that the analysis of the Schr\"odinger-Newton equation for the time evolution of a superposition of two Gaussian wave-packets, as in a two-slit experiment, demonstrates self-gravity interaction has a distinct effect on quantum interference. Interference for small mass particles  is virtually indistinguishable from that governed by pure Schr\"odinger evolution. For larger mass particles, quantum interference is suppressed. For intermediate mass values, interference with a reduced visibility is seen. Now in an actual experiment, observation of interference with reduced visibility can also be attributed to environmental effects. However, the fringe width $w$ emerges as a key element in distinguishing self-gravity effects from those of decoherence. It yields a definite signature of the effect of self-gravity as mass increases,  and should be verifiable experimentally, if matter wave interferometry experiments can be carried out at the appropriate length and mass scales. 

There appears to be a competition between the spread of the wave-function due to quantum evolution and its contraction due to self-gravity. 
It is clear that within this model, a superposition of two disjoint wave-packets will not spontaneously ``collapse" onto one of the two parts. One may have to consider interaction of the particle with an external localized body of larger mass, to see if it triggers a collapse. Such suggestions have been made in earlier works too \cite{kumar2000single}.

The deviation of $w$ vs $1/m$ from a straight line for large mass is expected, as there is a mass-dependent self-interaction potential affecting the dynamics of the particle.
If this phenomenon is experimentally corroborated, then there would be reason for further analysis of the origin and effects of the S-N potential in the Schr\"odinger equation. We believe our work provides sufficient reason for renewed experimental work in matter-wave interferometry, for larger mass particles. Apart from providing clues to the emergence of classicality from quantum mechanics, such experiments may also throw some light on the question as to whether a full quantum theory of gravity is needed, or semi-classical gravity is sufficient in several quantum mechanical contexts.

\section*{Acknowledgements}

This work was partially supported by the Department of Science and Technology, India through the grant DST/ICPS/QuST/Theme-3/2019/Q109.
SKS and RV would also like to thank Dr. Chandradew Sharma and Dr. Kinjal Banerjee for fruitful discussions related to the work.
TQ would like to thank Imtiyaz Ahmad Bhat for computational help.

\bibliographystyle{ieeetr}
\bibliography{Reference.bib}

\end{document}